\documentclass[12pt]{article}

\usepackage{graphicx}

\newcommand{\tr}{\mbox{Tr} \, }
\newcommand{\ket}[1]{\left | #1 \right \rangle}
\newcommand{\bra}[1]{\left \langle #1 \right |}

\newcommand{\proj}[1]{\ket{#1} \! \bra{#1}}

\newcommand{\ave}[1]{\left \langle #1 \right \rangle}
\newcommand{\superop}{{\cal E}}

\newcommand{\hilbert}{{\cal H}}
\newcommand{\relent}[2]{{\cal S}\left ( #1 || #2 \right )}
\newcommand{\supp}{\mbox{supp} \, }
\newcommand{\avail}{\mbox{$\cal A$}}
\newcommand{\bigsum}[1]{{\displaystyle \sum_{#1}}}
\newcommand{\Stherm}{S_{\theta}}
\newcommand{\zef}{\mbox{\sf zef} }
\newcommand{\lenobs}{\mbox{$\Lambda$} }

\begin{document}

\title{Relative entropy in quantum information theory}

\author{Benjamin Schumacher$^{(1)}$
        and Michael D. Westmoreland$^{(2)}$}
\maketitle
\begin{center}
{\sl
$^{(1)}$Department of Physics, Kenyon College, Gambier, OH 43022 USA \\
$^{(2)}$Department of Mathematical Sciences, Denison University,
 Granville, OH  43023 USA }
\end{center}

\section*{Abstract}

We review the properties of the quantum relative entropy function
and discuss its application to problems of classical and quantum
information transfer and to quantum data compression.
We then outline further uses of
relative entropy to quantify quantum entanglement and analyze
its manipulation.

\section{Quantum relative entropy}

In this paper we discuss several uses of
the quantum relative entropy function in quantum information theory.
Relative entropy methods have a number of advantages.  First of all,
the relative entropy functional satisfies some strong identities and
inequalities, providing a basis for good theorems.  Secondly,
the relative entropy has a natural interpretation in terms of
the statistical distinguishability of quantum states; closely
related to this is the picture of relative entropy as a ``distance''
measure between density operators.  These interpretations of the
relative entropy give insight about the meaning of the mathematical
constructions that use it.  Finally, relative entropy has found a
wide variety of applications in quantum information theory.

The usefulness of relative entropy in quantum information theory
should come as no surprise, since the classical relative entropy has
shown its power as a unifying concept in classical information theory
\cite{cover}.  Indeed, some of the results we will describe have
close analogues in the classical domain.  Nevertheless, the quantum
relative entropy can provide insights in contexts (such as the
quantification of quantum entanglement) that have no parallel in
classical ideas of information.

Let $Q$ be a quantum system described by a Hilbert space $\hilbert$.
(Throughout this paper, we will restrict our attention to systems with
Hilbert spaces having a finite number of dimensions.)  A pure state of
$Q$ can be described by a normalized vector $\ket{\psi}$ in $\hilbert$,
but a general (mixed) state requires a density operator $\rho$, which
is a positive semi-definite operator on $\hilbert$ with unit trace.
For the pure state $\ket{\psi}$, the density operator $\rho$ is simply
the projection operator $\proj{\psi}$;
otherwise, $\rho$ is a convex combination of projections.
The entropy $S(\rho)$ is defined to be
\begin{equation}
	S(\rho) = - \tr \rho \log \rho .
\end{equation}
The entropy is non-negative and equals zero if and only if $\rho$ is a pure
state.  (By ``$\log$'' we will mean a logarithm with base 2.)

Closely related to the entropy of a state is the relative entropy of a pair
of states.  Let $\rho$ and $\sigma$ be density operators, and define the
quantum relative entropy $\relent{\rho}{\sigma}$ to be
\begin{equation}
	\relent{\rho}{\sigma} = \tr \rho \log \rho - \tr \rho \log \sigma .
\end{equation}
(We read this as ``the relative entropy of $\rho$ with respect to $\sigma$''.)
This function has a number of useful properties:  \cite{wehrl}
\begin{enumerate}
	\item  $\relent{\rho}{\sigma} \geq 0$, with equality if and only if
		$\rho = \sigma$.
	\item  $\relent{\rho}{\sigma} < \infty$ if and only if $\supp \rho
		\subseteq \supp \sigma$.  (Here ``$\supp \rho$'' is the subspace
		spanned by eigenvectors of $\rho$ with non-zero eigenvalues.)
	\item  The relative entropy is continuous where it is not infinite.
	\item  The relative entropy is jointly convex in its arguments
		\cite{jointconvex}.
		That is, if $\rho_1$, $\rho_2$, $\sigma_1$ and $\sigma_2$ are
		density operators, and $p_1$ and $p_2$ are non-negative numbers
		that sum to unity (i.e., probabilities), then
		\begin{equation}
			\relent{\rho}{\sigma} \leq p_1  \relent{\rho_1}{\sigma_1}
					+ p_2  \relent{\rho_2}{\sigma_2}
		\end{equation}
		where $\rho = p_1 \rho_1 + p_2 \rho_2$ and $\sigma = p_1 \sigma_1
		+ p_2 \sigma_2$.  Joint convexity automatically implies convexity
		in each argument, so that (for example)
		\begin{equation}
		  \relent{\rho}{\sigma} \leq p_1  \relent{\rho_1}{\sigma}
			+ p_2  \relent{\rho_2}{\sigma} .
		\end{equation}
\end{enumerate}
The properties, especially property (1), motivate us to think of the
relative entropy as a kind of ``distance'' between density operators.
The relative entropy, which is not symmetric and which lacks a triangle
inequality, is not technically a metric; but it is a positive definite
directed measure of the separation of two density operators.

Suppose the density operator $\rho_k$ occurs with
probability $p_k$, yielding an average state $\rho = \bigsum{k} p_k \rho_k$,
and suppose $\sigma$ is some other density operator.  Then
\begin{eqnarray}
\sum_{k} p_{k} \relent{\rho_k}{\sigma}
	& = & \sum_{k} p_k \left ( \tr \rho_k \log \rho_k -
			\tr \rho_k \log \sigma \right ) \nonumber \\
	& = & \sum_{k} p_k  \left ( \tr \rho_k \log \rho_k - \tr \rho_k \log \rho +
			\tr \rho_k \log \rho - \tr \rho_k \log \sigma \right ) \nonumber \\
	& = & \sum_{k} p_k \left ( \tr \rho_k \log \rho_k - \tr \rho_k \log \rho
		\right ) + \tr \rho \log \rho - \tr \rho \log \sigma \nonumber \\
\sum_{k} p_{k} \relent{\rho_k}{\sigma}
	& = &  \sum_{k} p_{k} \relent{\rho_k}{\rho} + \relent{\rho}{\sigma} .
		\label{donaldident}
\end{eqnarray}
Equation~\ref{donaldident} is known as Donald's identity.  \cite{donald}

The classical relative entropy of two probability distributions
is related to the probability of distinguishing the two
distributions after a large but finite number of independent samples.
This is called Sanov's theorem \cite{cover}, and this result
has quantum analogue \cite{qsanov}.
Suppose $\rho$ and $\sigma$ are two possible states of the
quantum system $Q$, and suppose we are provided with $N$ identically
prepared copies of $Q$.  A measurement is made to determine whether the
prepared state is $\rho$, and the probability $P_N$ that the state $\sigma$
passes this test---in other words, is confused with $\rho$---is
\begin{equation}
	P_N \approx 2^{-N \relent{\rho}{\sigma}}
\end{equation}
as $N \rightarrow \infty$.  (We have assumed that the measurement made
is an optimal one for the purpose, and it is possible to show that an
asymptotically optimal measurement strategy can be found that depends
on $\rho$ but not $\sigma$.)

The quantum version of Sanov's theorem tells us that the quantum relative
entropy governs the asymptotic distinguishability of one quantum state from
another by means of measurements.  This further supports the view of
$\relent{\cdot}{\cdot}$ as a measure of ``distance''; two states are ``close''
if they are difficult to distinguish, but ``far apart'' if the probability
of confusing them is small.

The remainder of this paper is organized as follows.
Sections 2--5 apply relative entropy methods to the problem
of sending classical information by means of a (possibly
noisy) quantum channel.  Sections 6--7 consider the
transmission and compression of quantum information.
Sections 8--9 then apply relative entropy methods to the discussion
of quantum entanglement and its manipulation by local operations
and classical communication.  We conclude with a few remarks in
Section 10.

\section{Classical communication via quantum channels}

One of the oldest problems in quantum information theory is that of
sending classical information via quantum channels.
A sender (``Alice'') wishes to transmit classical information
to a receiver (``Bob'') using a quantum system as a communication
channel.  Alice will represent the message $a$, which occurs
with probability $p_a$, by preparing the channel in the
``signal state'' represented by the density operator $\rho_a$.
The average state of the channel will thus be $\rho = \bigsum{a}
p_a \rho_a$.
Bob will attempt to recover the message by making
a measurement of some ``decoding observable'' on the channel
system.

The states $\rho_a$ should be understood
here as the ``output'' states of the channel,
the states that Bob will attempt to
distinguish in his measurement.  In other words, the states
$\rho_a$ already include the effects of the dynamical evolution
of the channel (including noise) on its
way from sender to receiver.  The dynamics of the channel
will be described by a trace-preserving, completely positive
map $\superop$ on density operators \cite{cpmaps}.
The effect of $\superop$
is simply to restrict the set of output channel states
that Alice can arrange for Bob to receive.
If $\cal D$ is the set of all density operators, then
Alice's efforts can only produce output states in the
set $\avail = \superop ( {\cal D} )$, a convex, compact
set of density operators.

Bob's decoding observable is represented by a set of
positive operators $E_b$ such that $\bigsum{b} E_b = 1$.
If Bob makes his measurement on the state $\rho_a$,
then the conditional probability of measurement
outcome $b$ is
\begin{equation}
	P(b|a) = \tr \rho_a E_b .
\end{equation}
This yields a joint distribution over Alice's input
messages $a$ and Bob's decoded messages $b$:
\begin{equation}
	P(a,b) = p_a P(b|a) .
\end{equation}
Once a joint probability distribution exists between
the input and output messages (random variables
$A$ and $B$, respectively), the information transfer
can be analyzed by classical information theory.
The information obtained by Bob is given by the
{\em mutual information} $I(A:B)$:
\begin{equation}
	I(A:B) = H(A) + H(B) - H(A,B) \label{mutual}
\end{equation}
where $H$ is the Shannon entropy function
\begin{equation}
	H(X) = - \sum_{x} p(x) \log p(x) .
\end{equation}
Shannon showed that, if the channel is used many
times with suitable error-correcting codes,
then any amount of information
up to $I(A:B)$ bits (per use of the channel)
can be sent from Alice to Bob with
arbitrarily low probability of error \cite{cover}.
The classical capacity of the channel is
$C = \max I(A:B)$, where the maximum is taken
over all input probability distributions.  $C$ is
thus the maximum amount of information that may be
reliably conveyed per use of the channel.

In the quantum mechanical situation, for a given ensemble
of signal states $\rho_a$, Bob has many different choices
for his decoding observable.  Unless the signal states
happen to be orthogonal, no choice of observable will
allow Bob to distinguish perfectly between them.
A theorem stated by Gordon\cite{gordon} and
Levitin\cite{levitin} and first proved by Holevo\cite{holevo}
states that the amount of information accessible to Bob is
limited by $ I(A:B) \leq \chi $, where
\begin{equation}
	\chi = S(\rho) - \sum_{a} p_a S(\rho_a) .  \label{chidef}
\end{equation}
The quantity $\chi$ is non-negative, since the entropy
$S$ is concave.

More recently, Holevo \cite{holevo2} and
Schumacher and Westmoreland \cite{noisy}
have shown that this upper bound on $I(A:B)$ is
asymptotically achievable.  If Alice uses the same channel many
times and prepares long codewords of signal states, and
Bob uses an entangled decoding observable to distinguish these
codewords, then Alice can convey to Bob up to
$\chi$ bits of information per use of the channel,
with arbitrarily low probability of error.  (This fact was
established for pure state signals $\rho_a = \proj{\psi_a}$
in \cite{hjsww}.  In this case, $\chi = S(\rho)$.)

The Holevo bound $\chi$ can be expressed in terms of the
relative entropy:
\begin{eqnarray}
  \chi  & = &	- \tr \rho \log \rho
		+ \sum_a p_a \tr \rho_a \log \rho_a \nonumber \\
	& = &	\sum_a p_a \left ( \tr \rho_a \log \rho_a
			- \tr \rho_a \log \rho \right ) \nonumber \\
  \chi	& = &	\sum_a p_a \relent{\rho_a}{\rho} .
\end{eqnarray}
In geometric terms, $\chi$ is the average relative entropy
``directed distance'' from the average state $\rho$ to the
members of the signal ensemble.

Donald's identity (Equation~\ref{donaldident}) has a particularly simple
form in terms of $\chi$.  Given an ensemble and an additional state
$\sigma$,
\begin{equation}
	\sum_a p_a \relent{\rho_a}{\sigma} = \chi + \relent{\rho}{\sigma}.
\end{equation}
This implies, among other things, that
\begin{equation}
	\chi \leq \sum_a p_a \relent{\rho_a}{\sigma}  \label{chibound}
\end{equation}
with equality if and only if $\sigma = \rho$, the ensemble average
state.

\section{Thermodynamic cost of communication}

In this section and the next, we focus on
the transfer of classical information by means of a
quantum channel.

Imagine a student who attends college far from home
\cite{benthesis}.
Naturally, the student's family wants to know that the
student is passing his classes, and so they want the
student to report to them frequently over the telephone.
But the student is poor and cannot affort very many
long-distance telephone calls.
So they make the following arrangement:
each evening at the same time, the poor student will call home
only if he is failing one or more of this classes.
Otherwise, he will save the phone charges by {\em not}
calling home.

Every evening that the poor student does not call, therefore,
the family is receiving a message {\em via the telephone}
that his grades are good.  (That the telephone is being
used for this message can be seen from the fact that,
if the phone lines are knocked out for some reason,
the family can no longer make any inference from the
absence of a phone call.)

For simplicity, imagine that the student's grades
on successive days are independent and that the
probability that the student will be failing on a
given evening is $p$.  Then the information
conveyed each evening by the presence or absence
of a phone call is
\begin{equation}
	H(p) = - p \log p - (1-p) \log (1-p) .
\end{equation}
The cost of making a phone call is $c$,
while not making a phone call is free.  Thus, the
student's average phone charge is $cp$ per evening.
The number of bits of information per unit cost
is thus
\begin{equation}
\frac{H(p)}{cp} = \frac{1}{c} \left ( - \log p -
		\left ( \frac{1}{p} - 1 \right ) \log (1 - p) \right ) .
\end{equation}
If the poor student is very successful in his studies,
so that $p \rightarrow 0$,
then this ratio becomes unboundedly large, even though both
$H(p) \rightarrow 0$ and $cp \rightarrow 0$.  That is, the
student is able to send an arbitrarily large number of bits
per unit cost.  There is no irreducible cost for sending one
bit of information over the telephone.

The key idea in the story of the poor student is that
one possible signal---no phone call at all---has no
cost to the student.  The student can exploit this fact to use
the channel in a cost-effective way, by using the zero-cost
signal almost all of the time.

Instead of a poor student using a telephone, we can consider an
analogous quantum mechanical problem.  Suppose that a sender can
manipulate a quantum channel to produce (for the receiver)
one of two possible states, $\rho_0$ or $\rho_1$.
The state $\rho_0$ can be produced
at ``zero cost'', while the state $\rho_1$ costs a finite amount
$c_1 > 0$ to produce.  In the signal ensemble, the signal state
$\rho_1$ is used with probability $\eta$ and
$\rho_0$ with probability
$1 - \eta$, leading to an average state
\begin{equation}
	\rho = (1-\eta) \rho_0 + \eta \rho_1 .
\end{equation}
The average cost of creating a signal is thus $c = \eta c_1$.
For this ensemble,
\begin{equation}
	\chi = (1-\eta) \relent{\rho_0}{\rho} + \eta \relent{\rho_1}{\rho} .
\end{equation}
As discussed in the previous section,
$\chi$ is an asymptotically achievable upper bound
for the information transfered by the channel.

An upper bound for $\chi$ can be obtained from Donald's identity.
Letting $\rho_0$ be the ``additional'' state,
\begin{equation}
  \chi  \leq
	(1-\eta) \relent{\rho_0}{\rho_0} + \eta \relent{\rho_1}{\rho_0}
    = \eta \relent{\rho_1}{\rho_0} .
\end{equation}
Combining this with a simple lower bound, we obtain
\begin{equation}
	\eta \relent{\rho_1}{\rho} \leq \chi
                 \leq \eta \relent{\rho_1}{\rho_0} .
\end{equation}
If we divide $\chi$ by the average cost, we find an asymptotically
achievable upper bound for the number of bits sent through the channel
per unit cost.  That is,
\begin{equation}
	\frac{\chi}{c} \leq \frac{1}{c_1} \relent{\rho_1}{\rho_0} .
\end{equation}
Furthermore, equality holds in the limit that $\eta \rightarrow 0$.
Thus,
\begin{equation}
	\sup \frac{\chi}{c} = \frac{1}{c_1} \relent{\rho_1}{\rho_0} .
\end{equation}
In short, the relative entropy ``distance'' between the signal state $\rho_1$
and the ``zero cost'' signal $\rho_0$ gives the largest possible number of
bits per unit cost that may be sent through the channel---the ``cost
effectiveness'' of the channel.  If the state $\rho_0$ is a pure state,
or if we can find a usable signal state $\rho_1$ whose support is not contained
in the support of $\rho_0$, then $\relent{\rho_1}{\rho_0} = \infty$ and the
cost effectiveness of the channel goes to infinity as $\eta \rightarrow 0$.
(This is parallel to the situation of the poor student, who can
make the ratio of ``bits transmitted'' to ``average cost''
arbitrarily large.)

What if there are many possible signal states $\rho_1$, $\rho_2$, etc.,
with positive costs $c_1$, $c_2$, and so on?  If we assign the probability
$\eta q_k$ to $\rho_k$ for $k = 1, 2, \ldots$ (where $\bigsum{k} q_k = 1$),
and use $\rho_0$ with probability $1-\eta$, then we obtain
\begin{equation}
  \eta \sum_k q_k \relent{\rho_k}{\rho}
	\leq \chi \leq
  \eta \sum_k q_k \relent{\rho_k}{\rho_0} .
\end{equation}
The average cost of the channel is $c = \eta \bigsum{k} q_k c_k$.
This means that
\begin{equation}
	\frac{\chi}{c} \leq
		\frac{\sum_k q_k \relent{\rho_k}{\rho_0}}{\sum_k q_k c_k} .
\end{equation}

We now note the following fact about real numbers.  Suppose $a_n, b_n > 0$
for all $n$.  Then
\begin{equation}
	\frac{\sum_n a_n}{\sum_n b_n} \leq \max_n \frac{a_n}{b_n} .
\end{equation}
This can be proven by letting $R = \max (a_n / b_n)$ and pointing out
that $a_n \leq R b_n$ for all $n$.  Then
\begin{eqnarray*}
	\sum_n a_n & \leq & R \sum_n b_n \\
	\frac{\sum_n a_n}{\sum_n b_n} & \leq & R.
\end{eqnarray*}
In our context, this implies that
\begin{equation}
	\frac{\sum_k q_k \relent{\rho_k}{\rho_0}}{\sum_k q_k c_k}
		\leq \max_k \frac{q_k \relent{\rho_k}{\rho_0}}{q_k c_k}
\end{equation}
and thus
\begin{equation}
	\frac{\chi}{c} \leq \max_k \frac{\relent{\rho_k}{\rho_0}}{c_k} .
\end{equation}
By using only the ``most efficient state'' (for which the maximum on
the right-hand side is achieved) and adopting the ``poor student''
strategy of $\eta \rightarrow 0$, we can show that
\begin{equation}
     \sup \frac{\chi}{c} = \max_k \frac{\relent{\rho_k}{\rho_0}}{c_k} .
\end{equation}

These general considerations of an abstract ``cost'' of creating
various signals have an especially elegant development if we consider the
thermodynamic cost of using the channel.  The thermodynamic entropy
$\Stherm$ is related to the information-theoretic
entropy $S(\rho)$ of the state $\rho$ of the system by
\begin{equation}
	\Stherm = k \ln 2 \, \, S(\rho) .
\end{equation}
The constant $k$ is Boltzmann's constant.
If our system has a Hamiltonian operator $H$, then the thermodynamic
energy $E$ of the state is the expectation of the Hamiltonian:
\begin{equation}
	E = \ave{H} = \tr \rho H .
\end{equation}
Let us suppose that we have access to a thermal reservoir at
temperature $T$.
Then the ``zero cost'' state $\rho_0$ is the thermal equilibrium state
\begin{equation}
	\rho_0 = \frac{1}{Z}e^{-\beta H} ,
\end{equation}
where $\beta = 1 / kT$ and $Z = \tr e^{-\beta H}$.  ($Z$ is the
partition function.)

The free energy of the system in the presence of a thermal reservoir at
temperature $T$ is $F = E - T \Stherm$.  For the equilibrium
state $\rho_0$,
\begin{eqnarray}
	F_0  & = &  \tr \rho_0 H
		+ k T \ln 2 \left(- \log Z -
			\frac{\beta }{\ln 2} \tr \rho_0 H \right )
			\nonumber \\
	    & = &  - k T \ln 2 \, \log Z
\end{eqnarray}
The thermodynamic cost of the state $\rho_1$ is just the difference
$F_1 - F_0$ between the free energies of $\rho_1$ and the equilibrium
state $\rho_0$.  But this difference has a simple relation to the
relative entropy.  First, we note
\begin{equation}
	\tr \rho_1 \log \rho_0 = - \log Z - \beta \tr \rho_1 H ,
\end{equation}
from which it follows that \cite{lindblad}
\begin{eqnarray}
	F_1 - F_0 & = &
		\tr \rho_1 H + k T \ln 2 \, \tr \rho_1 \log \rho_1
			+ k T \ln 2 \, \log Z   \nonumber \\
	& = &  k T \ln 2 \left ( \tr \rho_1 \log \rho_1 -
			\tr \rho_1 \log \rho_0 \right ) \nonumber \\
	F_1 - F_0 & = & k T \ln 2 \, \, \relent{\rho_1}{\rho_0} .
\end{eqnarray}
If we use the signal state $\rho_1$ with probability $\eta$, then
the average thermodynamic cost is $f = \eta (F_1 - F_0)$.  The
number of bits sent per unit free energy is therefore
\begin{equation}
	\frac{\chi}{f} \leq \eta \frac{\relent{\rho_1}{\rho_0}}{f}
		= \frac{1}{k T \ln 2} .
\end{equation}
The same bound holds for all choices of the state $\rho_1$, and
therefore for all ensembles of signal states.

We can approach this upper bound if we make $\eta$ small, so
that
\begin{equation}
	\sup \frac{\chi}{f} = \frac{1}{kT \, \ln 2}
\end{equation}
In short, for {\em any} coding and decoding scheme
that makes use of the quantum channel,
the maximum number of bits that can be sent
per unit free energy is just $(k T \ln 2)^{-1}$.
Phrased another way, the minimum free energy cost
per bit is $k T \ln 2$.

This analysis can shed some light on Landauer's principle
\cite{landauer},
which states that the minimum thermodynamic cost
of information erasure is $k T \ln 2$ per bit.  From this point
of view, information erasure is simply information transmission
into the environment, which requires the expenditure of an
irreducible amount of free energy.

\section{Optimal signal ensembles}

Now we consider $\chi$-maximizing ensembles of states
from a given set $\avail$ of available (output) states,
without regard to the ``cost'' of each state.  Our
discussion in Section 2 tells us that the $\chi$-maximizing
ensemble is the one to use if we wish to maximize
the classical information transfer from Alice to Bob
via the quantum channel.  Call an ensemble that
maximizes $\chi$ an ``optimal'' signal
ensemble, and denote the maximum value of $\chi$ by $\chi^{*}$.
(The results of this section are developed in more detail in
\cite{optimal}.)

The first question is, of course, whether an optimal ensemble
exists.  It is conceivable that, though there is a least upper
bound $\chi^{*}$ to the possible values of $\chi$, no
particular ensemble in $\avail$ achieves it.
(This would be similar to the results in the last section,
in which the optimal cost effectiveness of the channel is
only achieved in a limit.)
However, an optimal ensemble does exist.
Uhlmann \cite{uhlmann}
has proven a result that goes most of the way.  Suppose our
underlying Hilbert space $\hilbert$ has dimension $d$ and
the set $\avail$ of available states is convex and compact.
Then given a fixed average state $\rho$, there exists an ensemble
of at most $d^2$ signal states $\rho_{a}$ that achieves the
maximum value of $\chi$ for that particular $\rho$.  The
problem we are considering is to maximize $\chi$ over all choices
of $\rho$ in $\avail$.  Since Uhlmann has shown that each
$\rho$-fixed optimal ensemble need involve no more than
$d^2$ elements, we only need to maximize $\chi$ over
ensembles that contain $d^2$ or fewer members.  The
set of such ensembles is compact and $\chi$ is a continuous
function on this set, so $\chi$ achieves its maximum value
$\chi^{*}$ for some ensemble with at most $d^2$ elements.

Suppose that the state $\rho_a$ occurs with probability $p_a$ in
some ensemble, leading to the average state $\rho$
and a Holevo quantity $\chi$.  We will now consider how
$\chi$ changes if we modify the ensemble slightly.
In the modified ensemble, a new state $\omega$ occurs
with probability $\eta$ and the state $\rho_a$ occurs with
probability $(1-\eta)p_a$.  For the modified ensemble,
\begin{eqnarray}
  \rho'	& = &	\eta \omega + (1-\eta) \rho \\
  \chi'	& = &	\eta \relent{\omega}{\rho'}
			+ (1-\eta) \sum_{a} p_a \relent{\rho_a}{\rho'} .
\end{eqnarray}

We can apply Donald's identity to these ensembles in
two different ways.  First, we can take the original optimal
ensemble and treat $\rho'$ as the other state ($\sigma$
in Eq.~\ref{donaldident}), obtaining:
\begin{equation}
  \sum_a p_a \relent{\rho_a}{\rho'} =  \chi + \relent{\rho}{\rho'} .
\end{equation}
Substituting this expression into the expression for $\chi'$ yields:
\begin{eqnarray}
  \chi'  & = &	\eta \relent{\omega}{\rho'}
		+ (1-\eta) \left( \chi + \relent{\rho}{\rho'} \right )
			\nonumber \\
  \Delta \chi & = & \chi' - \chi \nonumber \\
		& = & \eta \left ( \relent{\omega}{\rho'} - \chi \right )
			+ \eta \relent{\rho}{\rho'}
\end{eqnarray}
Our second application of Donald's identity is to the modified
ensemble, taking the original average state $\rho$ to play
the role of the other state:
\begin{eqnarray}
  \eta \relent{\omega}{\rho} + (1 - \eta) \chi
	& = & \chi' + \relent{\rho'}{\rho}  \\
\Delta \chi & = & \eta \left ( \relent{\omega}{\rho} - \chi
		\right ) - \relent{\rho'}{\rho} .
\end{eqnarray}
Since the relative entropy is never negative, we can conclude that
\begin{equation}
  \eta \left (  \relent{\omega}{\rho'} - \chi  \right )
	\leq \Delta \chi \leq
  \eta \left (  \relent{\omega}{\rho} - \chi \right ) .
	\label{deltachi}
\end{equation}
This gives upper and lower bounds for the change in $\chi$ if
we mix in an additional state $\omega$ to our original
ensemble.  The bounds are ``tight'', since as $\eta \rightarrow 0$,
$\relent{\omega}{\rho'} \rightarrow \relent{\omega}{\rho}$.

Very similar bounds for $\Delta \chi$ apply if we make more
elaborate modifications of our original ensemble, involving
more than one additional signal state.  This is described in
\cite{optimal}.

We say that an ensemble has the {\em maximal distance property}
if and only if, for any $\omega$ in $\avail$,
\begin{equation}
	\relent{\omega}{\rho} \leq \chi ,
\end{equation}
where $\rho$ is the average state and $\chi$ is the Holevo
quantity for the ensemble.  This property gives an interesting
characterization of optimal ensembles:
\begin{quote}
	{\bf Theorem:}  {\em An ensemble is optimal if and only if
		it has the maximum distance property.}
\end{quote}
We give the essential ideas of the proof here; further details
can be found in \cite{optimal}.

Suppose our ensemble has the maximum distance property.  Then, if
we add the state $\omega$ with probability $\eta$, the change
$\Delta \chi$ satisfies
\begin{equation}
   \Delta \chi \leq \eta \left (  \relent{\omega}{\rho} - \chi \right )
		\leq 0 .
\end{equation}
In other words, we cannot increase $\chi$ by mixing in an additional
state.  Consideration of more general changes to the ensemble leads
to the same conclusion that $\Delta \chi \leq 0$.  Thus, the ensemble
must be optimal, and $\chi = \chi^*$.

Conversely, suppose that the ensemble is optimal (with $\chi = \chi^*$).
Could there be a state $\omega$ in $\avail$ such that
$\relent{\omega}{\rho} > \chi^*$?  If there were such an
$\omega$, then by choosing $\eta$ small enough we could make
$\relent{\omega}{\rho'} > \chi^*$, and so
\begin{equation}
  \Delta \chi \geq \eta \left (  \relent{\omega}{\rho'} - \chi^* \right )
		> 0 .
\end{equation}
But this contradicts the fact that, if the original ensemble is
optimal, $\Delta \chi \leq 0$ for any change in the ensemble.
Thus, no such $\omega$ exists and the optimal ensemble satisfies
the maximal distance property.

Two corollaries follow immediately from this theorem.
First, we note that the support of the average state $\rho$
of an optimal ensemble must contain the support of every state
$\omega$ in $\avail$.  Otherwise, the relative entropy
$\relent{\omega}{\rho} = \infty$, contradicting the maximal
distance property.  The fact that $\rho$ has the largest
support possible could be called the {\em maximal support
property} of an optimal ensemble.

Second, we recall that $\chi^*$ is just the average relative
entropy distance of the members of the optimal ensemble from the
average state $\rho$:
\begin{displaymath}
	\chi^* = \sum_a p_a \relent{\rho_a}{\rho} .
\end{displaymath}
Since $\relent{\rho_a}{\rho} \leq \chi^*$ for each $a$, it
follows that whenever $p_a > 0$ we must have
\begin{equation}
	\relent{\rho_a}{\rho} = \chi^* .
\end{equation}
We might call this the {\em equal distance property} of
an optimal ensemble.

We can now give an explicit formula for $\chi^*$
that does not optimize over ensembles, but only over
states in $\avail$.
From Equation~\ref{chibound}, for any state $\sigma$,
\begin{equation}
	\chi \leq \sum_{a} p_a \relent{\rho_a}{\sigma}
\end{equation}
and thus
\begin{equation}
	\chi \leq \max_{\omega} \relent{\omega}{\sigma}
\end{equation}
where the maximum is taken over all $\omega$ in $\avail$.
We apply this inequality to the optimal ensemble, finding
the lowest such upper bound for $\chi^*$:
\begin{equation}
	\chi^* \leq \min_{\sigma} \left ( \max_{\omega}
			\relent{\omega}{\sigma} \right ) .
\end{equation}
But since the optimal ensemble has the maximal distance
property, we know that
\begin{equation}
	\chi^* = \max_{\omega} \relent{\omega}{\rho}
\end{equation}
for the optimal average state $\rho$.  Therefore,
\begin{equation}
	\chi^* = \min_{\sigma} \left ( \max_{\omega}
			\relent{\omega}{\sigma} \right ) .
\end{equation}

\section{Additivity for quantum channels}

The quantity $\chi^*$ is an asymptotically achievable upper bound
to the amount of classical information that can be sent using
available states of the channel system $Q$.  It is therefore tempting
to identify $\chi^*$ as the classical capacity of the quantum
channel.  But there is a subtlety here, which involves an important
unsolved problem of quantum information theory.

Specifically, suppose that two quantum systems $A$ and $B$
are available for use as communication channels.  The two systems
evolve independently according the product map $\superop^A \otimes
\superop^B$.  Each system can be considered as a separate
channel, or the joint system $AB$ can be analyzed as a single
channel.  It is not known whether the following holds
in general:
\begin{equation}
	\chi^{AB*} \stackrel{?}{=} \chi^{A*} + \chi^{B*} .
		\label{additivity}
\end{equation}
Since separate signal ensembles for $A$ and $B$ can be combined into
a product ensemble for $AB$, it is clear that
$\chi^{AB*} \geq \chi^{A*} + \chi^{B*}$.  However, the joint
system $AB$ also has other possible signal ensembles that use
entangled input states and that might perhaps have a Holevo bound
for the output states greater than $\chi^{A*} + \chi^{B*}$.

Equation~\ref{additivity} is the ``additivity conjecture'' for
the classical capacity of a quantum channel.  If the conjecture
is false, then the use of entangled input states
would sometimes increase the amount of classical information
that can be sent over two or more independent channels.
The classical capacity of a channel (which is defined
asymptotically, using many instances of the same channel)
would thus be greater than $\chi^*$ for a single instance of
a channel.  On the other hand, if the conjecture holds,
then $\chi^*$ is the classical capacity of the quantum channel.

Numerical calculations to date \cite{smolin_sim} support
the additivity conjecture for a variety of channels.
Recent work \cite{ruskai_king,holevo3} gives strong evidence that
Equation~\ref{additivity} holds for various special cases,
including channels described by unital maps.
We present here another partial result: $\chi^*$ is additive for any
``half-noisy'' channel, that is, a dual channel that is represented
by an map of the form ${\cal I}^A \otimes \superop^B$, where
${\cal I}^A$ is the identity map on $A$.

Suppose the joint system $AB$ evolves according to the map
${\cal I}^A \otimes \superop^B$, and let $\rho^A$ and $\rho^B$
be the average output states of optimal signal ensembles for
$A$ and $B$ individually.  We will show that the product
ensemble (with average state $\rho^A \otimes \rho^B$)
is optimal by showing that this ensemble
has the maximal distance property.  That is, suppose we
have another, possibly entangled input state of $AB$ that
leads to the output state $\omega^{AB}$.  Our aim is to
prove that $\relent{\omega^{AB}}{\rho^A \otimes \rho^B}
\leq \chi^{A*} + \chi^{B*}$.  From the definition of
$\relent{\cdot}{\cdot}$ we can show that
\begin{eqnarray}
\relent{\omega^{AB}}{\rho^A \otimes \rho^B} & = &
	- S \left ( \omega^{AB} \right ) - \tr \omega^A \log \rho^A
	- \tr \omega^B \log \rho^B   \nonumber \\
	& = & S \left ( \omega^A \right ) + S \left ( \omega^B \right )
		- S \left ( \omega^{AB} \right ) \nonumber \\
	& & \, \,+\, \relent{\omega^A}{\rho^A} + \relent{\omega^B}{\rho^B} .
	\label{relentprod}
\end{eqnarray}
(The right-hand expression has an interesting structure;
$S(\omega^A) + S(\omega^B) - S(\omega^{AB})$ is clearly
analogous to the mutual information defined in Equation~\ref{mutual}.)

Since $A$ evolves according to the identity map ${\cal I}^A$,
it is easy to see that $\chi^{A*} = d = \dim \hilbert^A$ and
\begin{equation}
	\rho^A = \left ( \frac{1}{d} \right ) \, 1^A .
\end{equation}
From this it follows that
\begin{equation}
	S \left ( \omega^A \right ) + \relent{\omega^A}{\rho^A}
		= \log d = \chi^{A*}  \label{Aresult}
\end{equation}
for any $\omega^A$.  This accounts for two of the terms on the
right-hand side of Equation~\ref{relentprod}.  The remaining
three terms require a more involved analysis.

The final joint state $\omega^{AB}$ is a mixed state, but we can always
introduce a third system $C$ that ``purifies'' the state.  That is,
we can find $\ket{\Omega^{ABC}}$ such that
\begin{equation}
	\omega^{AB} = \tr_C \proj{\Omega^{ABC}} .
\end{equation}
Since the overall state of $ABC$ is a pure state,
$S(\omega^{AB}) = S(\omega^C)$,
where $\omega^C$ is the state obtained by partial trace over $A$ and $B$.
Furthermore, imagine that a complete measurement is made on $A$, with the
outcome $k$ occuring with probability $p_k$.  For a given measurement
outcome $k$, the subsequent state of the remaining system $BC$ will
be $\ket{\Omega_k^{BC}}$.  Letting
\begin{eqnarray}
	\omega_k^B & = & \tr_C \proj{\Omega_k^{BC}} \nonumber \\
	\omega_k^C & = & \tr_B \proj{\Omega_k^{BC}} ,
\end{eqnarray}
we have that $S(\omega_k^B) = S(\omega_k^C)$ for all $k$.
Furthermore, by locality,
\begin{eqnarray}
	\omega^B = \sum_k p_k \omega_k^B \nonumber  \\
	\omega^C = \sum_k p_k \omega_k^C .
\end{eqnarray}
In other words, we have written both $\omega^B$ and $\omega^C$ as
ensembles of states.

We can apply this to get an upper bound on the remaining
terms in Equation~\ref{relentprod}
\begin{eqnarray}
  \lefteqn{S \left ( \omega^B \right )
	- S \left ( \omega^{AB} \right )
	+ \relent{\omega^B}{\rho^B}} \, \, \, \, \nonumber \\
  & = & S \left ( \omega^B \right )
		- \sum_k p_k S \left ( \omega_k^B \right ) \nonumber \\
  &  &  - S \left ( \omega^{C} \right )
		+ \sum_k p_k S \left ( \omega_k^C \right )
	+ \relent{\omega^B}{\rho^B} \nonumber \\
  & \leq &  \chi_{\omega}^B + \relent{\omega^B}{\rho^B} ,
\end{eqnarray}
where $\chi_{\omega}^B$ is the Holevo quantity for the ensemble
of $\omega_k^B$ states.  Donald's identity permits us to write
\begin{equation}
S \left ( \omega^B \right )
	- S \left ( \omega^{AB} \right )
	+ \relent{\omega^B}{\rho^B}
	= \sum_k p_k \relent{\omega_k^B}{\rho^B} .
	\label{Bresult}
\end{equation}

The $B$ states $\omega_k^B$ are all available output states
of the $B$ channel.  These states are obtained by making a complete
measurement on system $A$ when the joint system $AB$ is in the state
$\omega^{AB}$.  But this state was obtained from some initial $AB$ state
and a dynamical map ${\cal I}^A \otimes \superop^B$.  This map
commutes with the measurement operation on $A$ alone, so we could
equally well make the measurement {\em before} the action of
${\cal I}^A \otimes \superop^B$.  The $A$-measurement outcome $k$
would then determine the input state of $B$, which would evolve into
$\omega_k^B$.  Thus, for each $k$, $\omega_k^B$ is a possible
output of the $\superop^B$ map.

Since $\rho^B$ has the maximum distance property and the states
$\omega_k^B$ are available outputs of the channel,
$\relent{\omega_k^B}{\rho^B} \leq \chi^{B*}$ for every $k$.
Combining Equations~\ref{relentprod}, \ref{Aresult} and \ref{Bresult},
we find the desired inequality:
\begin{equation}
	\relent{\omega^{AB}}{\rho^A \otimes \rho^B} \leq
		\chi^{A*} + \chi^{B*} .
\end{equation}
This demonstrates that the product of optimal ensembles for $A$
and $B$ also has the maximum distance property for the possible
outputs of the joint channel, and so this product ensemble must
be optimal.  It follows that $\chi^{AB*} = \chi^{A*} + \chi^{B*}$
in this case.

Our result has been phrased for the case in which $A$ undergoes
``trivial'' dynamics ${\cal I}^A$, but the proof also
works without modification if the time evolution of
$A$ is unitary---that is, $A$ experiences ``distortion''
but not ``noise''.  If only one
of the two systems is noisy, then $\chi^*$ is additive.

The additivity conjecture for $\chi^*$ is closely related to
another additivity conjecture, the ``minimum output entropy''
conjecture \cite{ruskai_king,holevo3}.
Suppose $A$ and $B$ are systems with independent
evolution described by $\superop^A \otimes \superop^B$, and
let $\rho^AB$ be an output state of the channel with minimal
entropy $S(\rho^AB)$.  Is $\rho^{AB}$ a product state
$\rho^A \otimes \rho^B$?  The answer is not known in general;
but it is quite easy to show this in the half-noisy case
that we consider here.

\section{Maximizing coherent information}

When we turn from the transmission of classical information to
the transmission of quantum information, it will be helpful
to adopt an explicit description of the channel dynamics,
instead of merely specifying the set of available output
states $\avail$.
Suppose the quantum system $Q$ undergoes a dynamical evolution
described by the map $\superop$.  Since $\superop$ is a trace-preserving,
completely positive map, we can always find a representation of $\superop$
as a unitary evolution of a larger system \cite{cpmaps}.
In this representation, we imagine that an additonal ``environment''
system $E$ is present, initially in a pure state
$\ket{\breve{0}^E}$, and that
$Q$ and $E$ interact via the unitary evolution operator $U^{QE}$.
That is,
\begin{equation}
	\rho^Q = \superop(\breve{\rho}^Q)
		= \tr_E U^{QE} \left ( \breve{\rho}^Q \otimes
		\proj{\breve{0}^E} \right ) U^{QE\dagger} .
\end{equation}
For convenience, we denote an initial state of a system by the
breve accent (as in $\breve{\rho}^Q$), and omit
this symbol for final states.

The problem of sending quantum information through our channel can be
viewed in one of two ways:
\begin{enumerate}
\item  An unknown pure quantum state of $Q$ is to be transmitted.
	In this case, our criterion of success is the {\em average
	fidelity} $\bar{F}$, defined as follows.  Suppose the input
	state $\ket{\breve{\phi}_k}$ occurs with probability $p_k$ and
	leads to the output state $\rho_k$.  Then
	\begin{equation}
		\bar{F} = \sum_k p_k \bra{\breve{\phi}_k}
			\rho_k \ket{\breve{\phi}_k} .
	\end{equation}
	In general, $\bar{F}$ depends not only on the average input
	state $\breve{\rho}^Q$ but also on the particular pure
	state input ensemble.  \cite{qcoding}
\item  A second ``bystander'' system $R$ is present,
	and the joint system $RQ$ is
	initially in a pure entangled state $\ket{\breve{\Psi}^{RQ}}$.
	The system $R$ has ``trivial'' dynamics described by
	the identity map $\cal I$, so that the joint system evolves
	according to ${\cal I} \otimes \superop$, yielding a final
	state $\rho^{RQ}$.  Success is determined in this case by
	the {\em entanglement fidelity} $F_e$, defined by
	\begin{equation}
		F_e = \bra{\breve{\Psi}^{RQ}}
			\rho^{RQ} \ket{\breve{\Psi}^{RQ}} .
	\end{equation}
	It turns out, surprisingly, that $F_e$ is only dependent
	on $\superop$ and the input state $\breve{\rho}^{Q}$ of $Q$ alone.
	That is, $F_e$ is an ``intrinsic'' property of $Q$ and its
	dynamics.  \cite{entex}
\end{enumerate}
These two pictures of quantum information transfer are
essentially equivalent, since $F_e$ approaches unity if and only if
$\bar{F}$ approaches unity for every ensemble with the same
average input state $\breve{\rho}^Q$.
For now we adopt the second point of view, in which
the transfer of quantum information is essentially the transfer
of quantum entanglement (with the bystander system $R$) through
the channel.

The quantum capacity of a channel should be defined as
the amount of entanglement that can be transmitted through
the channel with $F_e \rightarrow 1$, if we allow ourselves to
use the channel many times and employ quantum error correction
schemes \cite{barnumetal}.
At present it is not known how to calculate this
{\em asymptotic} capacity of the channel in terms of the
properties of a single instance of the channel.

Nevertheless, we can identify some quantities that are useful
in describing the quantum information conveyed by the channel
\cite{coherent}.
A key quantity is the {\em coherent information} $I^Q$,
defined by
\begin{equation}
	I^{Q} = S \left( \rho^{Q} \right )
		- S \left ( \rho^{RQ} \right ) .
\end{equation}
This quantity is a measure of the final entanglement between $R$ and $Q$.
(The initial entanglement is measured by the entropy $S(\breve{\rho}^Q)$
of the initial state of $Q$, which of course equals $S(\breve{\rho}^R)$.
See Section  7 below.)
If we adopt a unitary representation for $\superop$, then the overall
system $RQE$ including the environment remains in a pure state from
beginning to end, and so $S(\rho^{RQ}) = S(\rho^{E})$.  Thus,
\begin{equation}
	I^{Q} = S \left( \rho^{Q} \right )
		- S \left ( \rho^{E} \right ) .
\end{equation}
Despite the apparent dependence of $I^{Q}$ on the systems $R$ and $E$,
it is in fact a function only of the map $\superop$ and the initial state
$\breve{\rho}^Q$ of $Q$.  Like the entanglement fidelity $F_e$, it is
an ``intrinsic'' characteristic of the channel system $Q$ and its
dynamics.

It can be shown that the coherent information $I^Q$ does not
increase if the map $\superop$ is followed by a
second independent map $\superop'$, giving an overall dynamics
described by $\superop' \circ \superop$.  That is, the coherent
information cannot be increased by any ``quantum data processing''
on the channel outputs.  The coherent information
is also closely related to quantum error correction.  Perfect
quantum error correction---resulting in $F_e = 1$ for the
final state---is possible if and only if the channel loses
no coherent information, so that $I^Q = S(\breve{\rho}^Q)$.
These and other properties lead us to consider $I^Q$ as a
good measure of the quantum information that is transmitted
through the channel \cite{coherent}.

The coherent information has an intriguing relation to the
Holevo quantity $\chi$, and thus to classical information
transfer (and to relative entropy) \cite{privacy}.
Suppose we describe
that the input state $\breve{\rho}^Q$ by an ensemble
of pure states $\ket{\breve{\phi}_k^Q}$:
\begin{equation}
	\breve{\rho}^Q = \sum_k p_k \proj{\breve{\phi}_k^Q} .
\end{equation}
We adopt a unitary representation for the evolution and note
that the initial pure state $\ket{\breve{\phi}_k^Q} \otimes
\ket{\breve{0}^E}$ evolves into a pure, possibly entangled
state $\ket{\phi_k^{QE}}$.  Thus, for each $k$
the entropies of the final states of $Q$ and $E$ are equal:
\begin{equation}
	S \left ( \rho_k^Q \right ) = S \left ( \rho_k^E \right ) .
\end{equation}
It follows that
\begin{eqnarray}
  I^Q  	& = & S \left( \rho^{Q} \right )
		- S \left ( \rho^{E} \right ) \nonumber \\
	& = & S \left( \rho^{Q} \right )
		- \sum_k p_k S \left( \rho_k^{Q} \right )
		- S \left ( \rho^{E} \right )
		+ \sum_k p_k S \left( \rho_k^{E} \right ) \nonumber \\
  I^Q	& = &	\chi^{Q} - \chi^{E} .   \label{privacyident}
\end{eqnarray}
Remarkably, the difference $\chi^{Q} - \chi^{E}$ depends only on
$\superop$ and the average input state $\breve{\rho}^{Q}$, not
the details of the environment $E$ or the exact choice of pure
state input ensemble.

The quantities $\chi^Q$ and $\chi^E$ are related to the
classical information transfer to the output system $Q$ and
to the environment $E$, respectively.  Thus,
Equation~\ref{privacyident} relates the classical and
quantum information properties of the channel.
This relation has been used to analyze the privacy
of quantum cryptographic channels \cite{privacy}.
We will use it here to give a
relative entropy characterization of the
the input state $\breve{\rho}^Q$
that maximizes the coherent information of the channel.

Let us suppose that $\breve{\rho}^Q$ is an input state that
maximizes the coherent information $I^Q$.
If we change the input state to
\begin{equation}
    {\breve{\rho}^Q}\mbox{$'$} = (1-\eta) \breve{\rho}^Q
			+ \eta \breve{\omega}^Q,
\end{equation}
for some pure state $\breve{\omega}^Q$, we
produces some change $\Delta I^Q$ in the coherent
information.  Viewing $\breve{\rho}^Q$
as an ensemble of pure states, this
change amounts to a modification of that
ensemble; and such a modification leads to changes in
the output ensembles for both system $Q$ and system
$E$.  Thus,
\begin{equation}
	\Delta I^Q = \Delta \chi^Q - \Delta \chi^E .
\end{equation}
We can apply Equation~\ref{deltachi} to bound both $\Delta \chi^Q$
and $\Delta \chi^E$ and obtain a lower bound for $\Delta I^Q$:
\begin{eqnarray}
	\Delta I^Q & \geq &
	  \eta \left ( \relent{\omega^Q}{{\rho^Q}'} - \chi^Q \right )
	- \eta \left ( \relent{\omega^E}{\rho^E} - \chi^E \right )
		\nonumber \\
	\Delta I^Q & \geq &
	  \eta \left ( \relent{\omega^Q}{{\rho^Q}'}
	- \relent{\omega^E}{\rho^E} - I^Q \right ) .
\end{eqnarray}
Since we assume that $I^Q$ is maximized for the input
$\breve{\rho}^Q$, then $\Delta I^Q \leq 0$ when we modify
the input state.  This must be true for every value of
$\eta$ in the relation above.  Whenever
$\relent{\omega^Q}{\rho^Q}$ is finite, we can
conclude that
\begin{equation}
	\relent{\omega^Q}{{\rho^Q}} - \relent{\omega^E}{\rho^E}
		\leq I^Q .  \label{distdiff}
\end{equation}
This is analogous to the maximum distance property for
optimal signal ensembles, except that it is the
difference of two relative entropy distances that is
bounded above by the maximum of $I^Q$.

Let us write Equation~\ref{privacyident} in terms of relative
entropy, imagining that the input state $\breve{\rho}^Q$
is written in terms of an ensemble of pure states
$\ket{\breve{\phi}_k^Q}$:
\begin{equation}
  I^Q = \sum_k p_k \left ( \relent{\rho_k^Q}{\rho^Q}
	- \relent{\rho_k^E}{\rho^E} \right ) .
\end{equation}
Every input pure state $\ket{\breve{\phi}_k^Q}$ in the input
ensemble with $p_k > 0$ will be in the support of $\breve{\rho}^Q$,
and so Equation~\ref{distdiff} holds.  Therefore, we can
conclude that
\begin{equation}
   I^Q = \relent{\rho_k^Q}{\rho^Q} - \relent{\rho_k^E}{\rho^E}
\end{equation}
for every such state in the ensemble.  Furthermore, {\em any}
pure state in the support of $\breve{\rho}^Q$ is a member of
some pure state ensemble for $\breve{\rho}^Q$.

This permits us to draw a remarkable conclusion.  If
$\breve{\rho}^Q$ is the input state that maximizes the
coherent information $I^Q$ of the channel, then for any
pure state $\breve{\omega}^Q$ in the support of
$\breve{\rho}^Q$,
\begin{equation}
    I^Q = \relent{\omega^Q}{\rho^Q} - \relent{\omega^E}{\rho^E} .
\end{equation}
This result is roughly analogous to the equal distance
property for optimal signal ensembles.  Together with
Equation~\ref{distdiff}, it provides a strong characterization
of the state that maximizes coherent information.

The additivity problem for $\chi^*$ leads us to ask whether the
maximum of the coherent information is additive when
independent channels are combined.  In fact, there are examples
known where $\max I^{AB} > \max I^A + \max I^B$; in other
words, entanglement between independent channels can increase
the amount of coherent information that can be sent through
them \cite{shorsmolin}.
The asymptotic behavior of coherent information and its
precise connection to quantum channel capacities are questions
yet to be resolved.

\section{Indeterminate length quantum coding}

In the previous section we saw that the relative entropy can be
used to analyze the coherent information ``capacity'' of a quantum
channel.  Another issue in quantum information theory is
{\em quantum data compression} \cite{qcoding}, which seeks
to represent quantum information using the fewest number
of qubits.  In this section we will see that the relative
entropy describes the cost of suboptimal quantum data
compression.

One approach to classical data compression is to use
variable length codes, in which the codewords are
finite binary strings of various lengths \cite{cover}.
The best-known examples are the Huffman codes.
The Shannon entropy $H(X)$ of a random variable
$X$ is a lower bound to the average codeword
length in such codes, and for Huffman codes this average
codeword length can be made arbitrarily close to $H(X)$.
Thus, a Huffman code optimizes the use of a
communication resources (number of bits required)
in classical communication without noise.

There are analogous codes for the compression of quantum
information.  Since coherent superpositions of codewords
must be allowed as codewords, these are called {\em
indeterminate length} quantum codes \cite{schu_sfi94}.
A quantum analogue to Huffman coding was recently
described by Braunstein et al. \cite{qhuffman}  An account of
the theory of indeterminate length quantum codes,
including the quantum Kraft inequality and the
condensability condition (see below),
will be presented in a forthcoming paper \cite{qkraft}.
Here we will outline a few results and demonstrate
a connection to the relative entropy.

The key idea in constructing an indeterminate length
code is that the codewords themselves must
carry their own length information.
For a classical variable length code, this requirement
can be phrased in two ways.  A {\em uniquely decipherable}
code is one in which any string of $N$ codewords can
be correctly separated into its individual codewords,
while a {\em prefix-free} code is one in which no codeword
is an initial segment of another codeword.  The
lengths of the codewords in each case satisfy the
Kraft-McMillan inequality:
\begin{equation}
	\sum_k 2^{-l_k} \leq 1 ,
\end{equation}
where is the sum is over the codewords and $l_k$ is the
length of the $k$th codeword.
Every prefix-free code is uniquely decipherable, so the
prefix-free property is a more restrictive property.
Nevertheless, it turns out that any uniquely decipherable
code can be replaced by a prefix-free code with the
same codeword lengths.

There are analogous conditions for indeterminate length
quantum codes, but these properties must be phrased
carefully because we allow coherent superpositions of codewords.
For example, a classical prefix-free
code is sometimes called an ``instantaneous'' code, since
as soon as a complete codeword arrives we can recognize
it at once and decipher it immediately.  However, if an
``instantaneous'' decoding procedure were to be attempted
for a quantum prefix-free code, it would destroy coherences
between codewords of different lengths.  Quantum codes
require that the entire string of codewords be deciphered
together.

The property of an indeterminate length quantum code that
is analogous to unique decipherability is called
{\em condensability}.  We digress briefly
to describe the condensability
condition.  We focus on {\em zero-extended forms} (\zef)
of our codewords.  That is, we cosider that our codewords
occupy an initial segment of a qubit register of
fixed length $n$, with $\ket{0}$'s following.  (Clearly
$n$ must be chosen large enough to contain the longest
codeword.)  The set of all \zef codewords spans
a subspace of the Hilbert space of register states.
We imagine that the output of a quantum information
source has been mapped unitarily to the \zef codeword
space of the register.  Our challenge is to take
$N$ such registers and ``pack'' them together in a
way that can exploit the fact that some of the codewords
are shorter than others.

If codeword states must carry their own length information,
there must be a {\em length observable} \lenobs on the
\zef codeword space with the following two properties:
\begin{itemize}
 \item  The eigenvalues of \lenobs are integers
	$1, \ldots , n$, where $n$ is the
	length of the register.
 \item  If $\ket{\psi_{\zef}}$ is an eigenstate of \lenobs
	with eigenvalue $l$, then it has the form
	\begin{equation}
		\ket{\psi_{\zef}} = \ket{\psi^{1 \cdots l}
		0^{l+1 \cdots n}} .
  	\end{equation}
	That is, the last $n - l$ qubits in the register are
	in the state $\ket{0}$ for a \zef codeword of length $l$.
\end{itemize}
For register states not in the \zef subspace, we can take
$\lenobs = \infty$.

A code is {\em condensable} if the following
condition holds:  For any $N$, there is a unitary
operator $U$ (depending on $N$) that maps
\begin{displaymath}
   \underbrace{\ket{\psi_{1,\zef}} \otimes \cdots \otimes
     \ket{\psi_{N,\zef}}}_{N n \, \mbox{\footnotesize qubits}}
     \rightarrow
   \underbrace{\ket{\Psi_{1 \cdots N}}}_{N n \, \mbox{\footnotesize qubits}}
\end{displaymath}
with the property that, if the individual codewords are all length
eigenstates, then $U$ maps the codewords to a \zef string of the
$N n$ qubits---that is, one with $\ket{0}$'s after the first
$L = l_{1} + \cdots + l_{N}$ qubits:
\begin{displaymath}
	\ket{\psi_{1}^{1 \cdots l_1} 0^{l_1 + 1 \cdots n}}
		\otimes \cdots \otimes
	\ket{\psi_{N}^{1 \cdots l_{N}} 0^{l_N + 1 \cdots n}}
		\rightarrow
	\ket{\Psi^{1 \cdots L} 0^{L + 1 \cdots Nn}} .
\end{displaymath}
The unitary operator $U$ thus ``packs'' $N$ codewords, given
in their \zef forms, into a ``condensed'' string that has
all of the trailing $\ket{0}$'s at the end.
The unitary character of the packing protocol automatically
yields an ``unpacking'' procedure given by $U^{-1}$.  Thus,
if the quantum code is condensable, a packed string of $N$
codewords can be coherently sorted out into separated \zef
codewords.

The quantum analogue of the Kraft-McMillan inequality states
that, for any indeterminate length quantum code that is
condensable, the length observable \lenobs on the subspace
of \zef codewords must satisfy
\begin{equation}
      \tr 2^{-\lenobs} \leq 1 , \label{qkraft}
\end{equation}
where we have restricted our trace to the \zef subspace.
We can construct a density operator $\omega$
(a positive operator of unit trace)
on the \zef subspace by letting
$K = \tr 2^{-\lenobs} \leq 1$ and
\begin{equation}
	\omega = \frac{1}{K} \, 2^{-\lenobs} .
\end{equation}
The density operator $\omega$ is generally not the same
as the actual density operator $\rho$ of the \zef
codewords produced by the quantum information source.
The average codeword length is
\begin{eqnarray}
	\bar{l} & = & \tr \rho \lenobs \nonumber \\
		& = & - \tr \rho \log \left ( 2^{-\lenobs} \right )
			\nonumber \\
		& = & - \tr \rho \log \omega - \log K \nonumber \\
	\bar{l} & = & S(\rho) + \relent{\rho}{\omega} - \log K .
		\label{barleq}
\end{eqnarray}
Since $\log K \leq 0$ and the relative entropy
is positive definite,
\begin{equation}
	\bar{l} \geq S(\rho) .  \label{vonNeumannbound}
\end{equation}
The average codeword length
must always be at least as great as the von Neuman
entropy of the information source.

Equality for Equation~\ref{vonNeumannbound}
can be approached asymptotically using block
coding and a quantum analogue of Huffman (or Shannon-Fano)
coding.  For special cases in which the eigenvalues of
$\rho$ are of the form $2^{-m}$, then a code exists
for which $\bar{l} = S(\rho)$, without the asymptotic
limit.  In either case, we say that a code satisfying
$\bar{l} = S(\rho)$ is a length optimizing quantum code.
Equation~\ref{barleq} tells us that, if we have a length
optimizing code, $K = 1$ and
\begin{equation}
	\rho = \omega = 2^{-\lenobs} .
\end{equation}
The condensed string of $N$ codewords has $Nn$ qubits,
but we can discard all but about $N \bar{l}$ of them
and still retain high fidelity.  That is, $\bar{l}$
is the asymptotic number of qubits that must be used
per codeword to represent the quantum information
faithfully.

Suppose that we have an indeterminate length quantum
code that is designed for the wrong density operator.
That is, our code is length optimizing for some other
density operator $\omega$, but $\rho \neq \omega$.
Then (recalling that $K = 1$ for a length optimizing
code, even if it is optimizing for the wrong density
operator),
\begin{equation}
	\bar{l} = S(\rho) + \relent{\rho}{\omega} .
\end{equation}
$S(\rho)$ tells us the number of qubits necessary to represent
the quantum information if we used a length optimizing
code for $\rho$.  (As we have mentioned, such codes
always exist in an asymptotic sense.)  However, to
achieve high fidelity in the situation where we have
used a code designed for $\omega$, we have to use
at least $\bar{l}$ qubits per codeword, an additional
cost of $\relent{\rho}{\omega}$ qubits per codeword.

This result gives us an interpretation of the relative
entropy function $\relent{\rho}{\omega}$ in terms
of the physical resources necessary to accomplish
some task---in this case, the additional cost (in qubits)
of representing the quantum information described by
$\rho$ using a coding scheme optimized for $\omega$.
This is entirely analogous to the situation for
classical codes and classical relative entropy
\cite{cover}.  A fuller development of this analysis
will appear in \cite{qkraft}.

\section{Relative entropy of entanglement}

One recent application of relative entropy has been to quantify
the entanglement of a mixed quantum state of two systems
\cite{relententangle}.  Suppose
Alice and Bob share a joint quantum system $AB$ in the state
$\rho^{AB}$.  This state is said to be {\em separable} if it is
a product state or else a probabilistic combination of product
states:
\begin{equation}
	\rho^{AB} = \sum_k p_k \rho_k^A \otimes \rho_k^B .
\end{equation}
Without loss of generality, we can if we wish
take the elements in this ensemble of product states
to be pure product states.  Systems in separable states
display statistical correlations having perfectly
ordinary ``classical'' properties---that is, they
do not violate any sort of Bell inequality.  A separable
state of $A$ and $B$ could also be created from scratch
by Alice and Bob using only local quantum operations
(on $A$ and $B$ separately) and the exchange of classical
information.

States which are not separable are said to be {\em entangled}.
These states cannot be made by local operations and classical
communication; in other words, their creation requires the
exchange of {\em quantum} information between Alice and Bob.
The characterization of entangled states and their possible
transformations has been a central issue in much recent work
on quantum information theory.

A key question is the quantification of entanglement, that is,
finding numerical measures of the entanglement of a quantum
state $\rho^{AB}$ that have useful properties.  If the joint
system $AB$ is in a pure state $\ket{\Psi^{AB}}$, so that the
subsystem states are
\begin{equation}
	\begin{array}{c}
		\rho^A = \tr_B \proj{\Psi^{AB}}  \\[.5ex]
		\rho^B = \tr_A \proj{\Psi^{AB}}
	\end{array}
\end{equation}
then the entropy $S(\rho^A) = S(\rho^B)$ can be used to measure
the entanglement of $A$ and $B$.  This measure has many appealing
properties.  It is zero if and only if $\ket{\Psi^{AB}}$ is
separable (and thus a product state).  For an ``EPR pair'' of
qubits---that is, a state of the general form
\begin{equation}
	\ket{\phi^{AB}} = \frac{1}{\sqrt{2}} \left (
				\ket{0^A 0^B} + \ket{1^A 1^B} \right ) ,
\end{equation}
the susbsystem entropy $S(\rho^A) = 1$ bit.

The subsystem entropy is also an asymptotic measure, both of the
resources necessary to create the particular entangled pure state,
and of the value of the state as a resource \cite{entconcentrate}.
That is, for sufficiently large $N$,
\begin{itemize}
\item  approximately $N S(\rho^A)$ EPR pairs are required to create
	$N$ copies of $\ket{\Psi^{AB}}$ by local operations and
	classical communication; and
\item  approximately $N S(\rho^A)$ EPR pairs can be created from
	$N$ copies of $\ket{\Psi^{AB}}$ by local operations and
	classical communication.
\end{itemize}

For mixed entangled states $\rho^{AB}$ of the joint system $AB$,
things are not so well-established.  Several different measures of
entanglement are known, including \cite{mammoth}
\begin{itemize}
\item the {\em entanglement of formation} $E(\rho^{AB})$, which is the
	minimum asymptotic number of EPR pairs required to create
	$\rho^{AB}$ by local operations and classical communication;
	and
\item the {distillable entanglement} $D(\rho^{AB})$, the maximum
	asymptotic number of EPR pairs that can be created from
	$\rho^{AB}$ by entanglement purification protocols involving
	local operations and classical communication.
\end{itemize}
Bennett et al. \cite{mammoth}
further distinguish $D_1$ and $D_2$, the distillable entanglements
with respect to purification protocols that allow one-way and two-way
classical communication, respectively.  All of these
measures reduce to the subsystem entropy $S(\rho^A)$
if $\rho^{AB}$ is a pure entangled state.

These entanglement measures are not all equal; furthermore, explicit
formulas for their calculation are not known in most cases.  This
motivates us to consider alternate measures of entanglement with
more tractable properties and which have useful relations to the
asymptotic measures $E$, $D_1$ and $D_2$.

A state $\rho^{AB}$ is entangled inasmuch as it is not a separable
state, so it makes sense to adopt as a measure of entanglement
a measure of the
distance of $\rho^{AB}$ from the set $\Sigma^{AB}$ of separable
states of $AB$.  Using relative entropy as our ``distance'',
we define the {\em relative entropy of entanglement} $E_r$ to be
\cite{relententangle}
\begin{equation}
	E_r \left ( \rho^{AB} \right ) =
		\min_{\sigma^{AB} \in \Sigma^{AB}}
		\relent{\rho^{AB}}{\sigma^{AB}} .
\end{equation}
The relative entropy of entanglement has several handy properties.
First of all, it reduces to the subsystem entropy $S(\rho^A)$ whenever
$\rho^{AB}$ is a pure state.  Second, suppose we write $\rho^{AB}$ as
an ensemble of pure states $\ket{\psi_k^{AB}}$.  Then
\begin{equation}
	E_r \left ( \rho^{AB} \right )
		\leq \sum_k p_k S \left ( \rho_k^A \right )
\end{equation}
where $\rho_k^A = \tr_B \proj{\psi_k^{AB}}$.  It follows from this
that $E_r \leq E$ for any state $\rho^{AB}$.

Even more importantly, the relative entropy of entanglement $E_r$ can be
shown to be non-increasing on average under local operations by Alice
and Bob together with classical communication between them.

The quantum version of Sanov's theorem gives the relative entropy of
entanglement an interpretation in terms of the statistical
distinguishability of $\rho^{AB}$ and the ``least distinguishable''
separable state $\sigma^{AB}$.
The relative entropy of entanglement is thus a useful and well-motivated
measure of the entanglement of a state $\rho^{AB}$ of a joint system,
both on its own terms and as a surrogate for less tractable
asymptotic measures.

\section{Manipulating multiparticle entanglement}

The analysis in this section closely follows that of
Linden et al. \cite{linden}, who provides a more
detailed discussion of the main result here and its applications.

Suppose Alice, Bob and Claire initially share three qubits in a ``GHZ
state''
\begin{equation}
	\ket{\Psi^{ABC}} = \frac{1}{\sqrt{2}} \left (
		\ket{0^A 0^B 0^C} + \ket{1^A 1^B 0^C} \right ) .
\end{equation}
The mixed state $\rho^{BC}$ shared by Bob and Claire is, in fact,
not entangled at all:
\begin{equation}
	\rho^{BC} = \frac{1}{2} \left (
			\proj{0^B 0^C} + \proj{1^B 1^C} \right ) .
\end{equation}
No local operations performed by Bob and Claire can produce an
entangled state from this starting point.  However, Alice can
create entanglement for Bob and Claire.  Alice
measures her qubit in the basis $\{ \ket{+^A},\ket{-^A} \}$,
where
\begin{equation}
	\ket{\pm^A} = \frac{1}{\sqrt{2}}
			\left ( \ket{0^A} \pm \ket{1^A} \right ).
\end{equation}
It is easy to verify that the state of Bob and Claire's qubits
after this measurement, depending on the measurement outcome,
must be one of the two states
\begin{equation}
	\ket{\phi^{BC}_{\pm}} = \frac{1}{\sqrt{2}} \left (
			\ket{0^A 0^B} \pm \ket{1^A 1^B} \right ) ,
\end{equation}
both of which are equivalent (up to a local unitary transformation
by either Bob or Claire) to an EPR pair.  In other words, if Alice
makes a local measurement on her qubit and then announces the
result by classical communication, the GHZ triple can be converted
into an EPR pair for Bob and Claire.

When considering the manipulation of quantum entanglement
shared among several parties, we must therefore bear in mind that
the entanglement between subsystems can both increase and
decrease, depending on the situation.  This raises several
questions:
Under what circumstances can Alice increase Bob and Claire's
entanglement?  How much can she do so?  Are there any costs
involved in the process?

To study these questions, we must give a more detailed account
of ``local operations and classical communication''.  It turns
out that Alice, Bob and Claire can realize any local operation
on their joint system $ABC$ by a combination of the following:
\begin{itemize}
	\item  Local unitary transformations on the subsystems
		$A$, $B$ and $C$;
	\item  Adjoining to a subsystem additional local ``ancilla''
		qubits in a standard state $\ket{0}$;
	\item  Local ideal measurements on the (augmented)
		subsystems $A$, $B$ and $C$; and
	\item  Discarding local ancilla qubits.
\end{itemize}
Strictly speaking, though, we do not need to include the last
item.  That is, any protocol that involves discarding ancilla
qubits can be replaced by one in which the ancillas are simply
``set aside''---not used in future steps, but not actually
gotten rid of.  In a similar vein, we can imagine that
the ancilla qubits required are already present in the subsystems
$A$, $B$ and $C$, so the second item in our list is redundant.
We therefore need to consider only local unitary transformations
and local ideal measurements.

What does classical communication add to this?  It is sufficient
to suppose that Alice, Bob and Claire have complete
information---that is, they are aware of all operations
and the outcomes of all measurements performed by each of them,
and thus know the global state of $ABC$ at every stage.  Any
protocol that involved an incomplete sharing of information
could be replaced by one with complete sharing, simply by ignoring
some of the messages that are exchanged.

Our local operations (local unitary transformations and
local ideal measurements) always take an initial pure state to a
final pure state.  That is, if $ABC$ starts in the joint state
$\ket{\Psi^{ABC}}$, then the final state will be a pure state
$\ket{\Psi_k^{ABC}}$ that depends on the joint outcome $k$ of all
the measurements performed.  Thus, $ABC$ is
always in a pure state known to all parties.

It is instructive to consider the effect of local operations
on the entropies of the various subsystems of $ABC$.  Local
unitary transformations leave $S(\rho^A)$, $S(\rho^B)$ and
$S(\rho^C)$ unchanged.  But suppose that Alice makes an ideal
measurement on her subsystem, obtaining outcome $k$ with
probability $p_k$.  The initial global state is $\ket{\Psi^{ABC}}$
and the final global state is $\ket{\Psi_k^{ABC}}$, depending
on $k$.  For the initial subsystem states, we have that
\begin{equation}
	S \left ( \rho^A \right ) = S \left ( \rho^{BC} \right )
\end{equation}
since the overall state is a pure state.  Similarly, the various
final subsystem states satisfy
\begin{equation}
	S \left ( \rho_k^A \right ) = S \left ( \rho_k^{BC} \right ) .
\end{equation}
But an operation on $A$ cannot change the average state of $BC$:
\begin{equation}
	\rho^{BC} = \sum_k p_k \rho_k^{BC} . \label{aveBCstate}
\end{equation}
Concavity of the entropy gives
\begin{equation}
	S \left ( \rho^{BC} \right ) \geq
         \sum_k p_k S \left ( \rho_k^{BC} \right )
\end{equation}
and therefore
\begin{equation}
	S \left ( \rho^{A} \right ) \geq
	\sum_k p_k S \left ( \rho_k^{A} \right ) .
\end{equation}
Concavity also tells us that
$S(\rho^B) \geq \bigsum{k} p_k S(\rho_k^B)$, etc.,
and similar results hold for local measurements
performed by Bob or Claire.

We now return to the question of how much Alice can
increase the entanglement shared by Bob and Claire.
Let us measure the bipartite entanglement of the
system $BC$ (which may be in a mixed state) by
the relative entropy of entanglement $E_r(\rho^{BC})$,
and let $\sigma^{BC}$ be the separable state of $BC$
for which
\begin{equation}
	E_r(\rho^{BC}) = \relent{\rho^{BC}}{\sigma^{BC}} .
\end{equation}
No local unitary operation can change $E_r(\rho^{BC})$;
furthermore, no local measurement by Bob or Claire can
increase $E_r(\rho^{BC})$ on average.  We need only
consider an ideal measurement performed by Alice
on system $A$.  Once again we suppose that outcome
$k$ of this measurement occurs with probability $p_k$,
and once again Equation~\ref{aveBCstate} holds.
Donald's identity tells us that
\begin{equation}
	\sum_k p_k \relent{\rho_k^{BC}}{\sigma^{BC}}
		= \sum_k p_k \relent{\rho_k^{BC}}{\rho^{BC}}
		+ \relent{\rho^{BC}}{\sigma^{BC}} .
\end{equation}
But $E_r(\rho_k^{BC}) \leq \relent{\rho_k^{BC}}{\sigma^{BC}}$ for
every $k$, leading to the following inequality:
\begin{equation}
	\sum_k p_k E_r(\rho_k^{BC}) - E_r(\rho^{BC})
		\leq \sum_k p_k \relent{\rho_k^{BC}}{\rho^{BC}} .
\end{equation}
We recognize the left-hand side of this inequality $\chi$ for
the ensemble of post-measurement states of $BC$, which we can
rewrite using the definition of $\chi$ in Equation~\ref{chidef}.
This yields:
\begin{eqnarray}
	\sum_k p_k E_r(\rho_k^{BC}) - E_r(\rho^{BC})
	  & \leq &  S \left( \rho^{BC} \right )
 	  - \sum_k p_k S \left ( \rho_k^{BC} \right ) \nonumber \\
	& = & S \left( \rho^{A} \right )
 	  - \sum_k p_k S \left ( \rho_k^{A} \right ) ,
\end{eqnarray}
since the overall state of $ABC$ is pure at every stage.

To summarize, in our model (in which all measurements are ideal,
all classical information is shared, and no classical or quantum
information is ever discarded), the following principles hold:
\begin{itemize}
\item  The entropy of any subsystem $A$ cannot be increased on average
	by any local operations.
\item  The relative entropy of entanglement of two subsystems
	$B$ and $C$ cannot be increased on average by local
	operations on those two subsystems.
\item  The relative entropy of entanglement of $B$ and $C$ can be
	increased by a measurement performed on a third subsystem $A$,
	but the average increase in $E_r^BC$ is no larger than the
	average decrease in the entropy of $A$.
\end{itemize}

We say that a joint state $\ket{\Psi_1^{ABC}}$ can be transformed
{\em reversibly} into $\ket{\Psi_2^{ABC}}$ if,
for sufficiently large $N$, $N$ copies of
$\ket{\Psi_1^{ABC}}$ can be transformed with high probability (via local
operations and classical communication) to approximately $N$ copies of
$\ket{\Psi_2^{ABC}}$, and {\em vice versa}.  The qualifiers in this
description are worth a comment or two.  ``High probability'' reflects
the fact that, since the local operations may involve measurements, the
actual final state may depend on the exact measurement outcomes.
``Approximately $N$ copies'' means more than $(1-\epsilon)N$ copies,
for some suitably small $\epsilon$ determined in advance.
We denote this reversibility relation by
\begin{displaymath}
	\ket{\Psi_1^{ABC}} \leftrightarrow \ket{\Psi_2^{ABC}} .
\end{displaymath}
Two states that are related in this way are essentially equivalent
as ``entanglement resources''.  In the large $N$ limit, they may
be interconverted with arbitrarily little loss.

Our results for entropy and relative entropy of entanglement
allow us to place necessary conditions on the reversible
manipulation of multiparticle entanglement.  For example, if
$\ket{\Psi_1^{ABC}} \leftrightarrow \ket{\Psi_2^{ABC}}$,
then the two states must have exactly the same subsystem
entropies.  Suppose instead that $S(\rho_1^A) < S(\rho_2^A)$.
Then the transformation of $N$ copies of $\ket{\Psi_1^{ABC}}$
into about $N$ copies of $\ket{\Psi_2^{ABC}}$ would involve
an increase in the entropy of subsystem $A$, which cannot happen
on average.

In a similar way, we can see that $\ket{\Psi_1^{ABC}}$ and
$\ket{\Psi_2^{ABC}}$ must have the same relative entropies of
entanglement for every pair of subsystems.  Suppose instead
that $E_{r,1}^{BC} < E_{r,2}^{BC}$.  Then the transformation
of $N$ copies of $\ket{\Psi_1^{ABC}}$ into about $N$ copies
of $\ket{\Psi_2^{ABC}}$ would require an increase in $E_r^{BC}$.
This can take place if a measurement is performed on $A$,
but as we have seen this would necessarily involve a decrease
in $S(\rho^A)$.  Therefore, reversible transformations of
multiparticle entanglement must preserve both subsystem
entropies and the entanglement (measured by $E_r$) of pairs
of subsystems.

As a simple example of this, suppose Alice, Bob and Claire share
two GHZ states.  Each subsystem has an entropy of 2.0 bits.
This would also be the case if Alice, Bob and Claire shared
three EPR pairs, one between each pair of participants.  Does
it follow that two GHZs can be transformed reversibly (in the
sense described above) into three EPRs?

No.  If the three parties share two GHZ triples, then Bob and
Claire are in a completely unentangled state, with $E_r^{BC} = 0$.
But in the ``three EPR'' situation, the relative entropy of
entanglement $E_r^{BC}$ is 1.0 bits, since they share an EPR
pair.  Thus, two GHZs cannot be reversibly transformed into
three EPRs; indeed, $2N$ GHZs are inequivalent to $3N$ EPRs.

Though we have phrased our results for three parties, they are
obviously applicable to situations with four or more separated
subsystems.  In reversible manipulations of multiparticle
entanglement, all subsystem entropies (including the entropies
of clusters of subsystems) must remain constant, as well as
the relative entropies of entanglement of all pairs of subsystems
(or clusters of subsystems).

\section{Remarks}

The applications discussed here show the power and the versatility
of relative entropy methods in attacking problems of quantum
information theory.  We have derived useful fundamental results
in classical and quantum information transfer, quantum data
compression, and the manipulation of quantum entanglement.
In particular, Donald's identity proves to be an extremely
useful tool for deriving important inequalities.

One of the insights provided by quantum information theory is that
the von Neumann entropy $S(\rho)$ has an interpretation (actually
several interpretations) as a measure of the resources necessary to
perform an information task.  We have seen that the relative entropy
also supports such interpretations.  We would especially like to
draw attention to the results in Sections 3 on the cost of
communication and Section 7 on quantum data compression, which are
presented here for the first time.

We expect that relative entropy techniques will be central to further
work in quantum information theory.  In particular, we think that
they show promise in resolving the many perplexing additivity problems
that face the theory at present.  Section 5, though not a very strong
result in itself, may point the way along this road.

The authors wish to acknowledge the invaluable help of many
colleagues.  T. Cover, M. Donald, M. Neilsen, M. Ruskai,
A. Uhlmann and V. Vedral have given us indispensible guidance
about the properties and meaning of the relative entropy function.
Our work on optimal signal ensembles and the additivity
problem was greatly assisted by conversations with
C. Fuchs, A. Holevo, J. Smolin, and W. Wootters.
Results described here on reversibility for transformations
of multiparticle entanglement were obtained in the course
of joint work with N. Linden and S. Popescu.  We would like to thank
the organizers of the AMS special session on ``Quantum Information
and Computation'' for a stimulating meeting and an opportunity to
pull together several related ideas into the present paper.  We
hope it will serve as a spur for the further application of relative
entropy methods to problems of quantum information theory.

\end{document}